\documentclass[12pt,preprint]{aastex}

\shorttitle{}

\shortauthors{Liu et al.}

\begin{document}

\title{Plasma and Magnetic Field Characteristics of Solar Coronal Mass 
Ejections in Relation to Geomagnetic Storm Intensity and Variability}

\author{Ying D. Liu\altaffilmark{1}, Huidong Hu\altaffilmark{1,2}, Rui Wang\altaffilmark{1},
Zhongwei Yang\altaffilmark{1}, Bei Zhu\altaffilmark{1,2}, Yi A. Liu\altaffilmark{1,2}, 
Janet G. Luhmann\altaffilmark{3}, and John D. Richardson\altaffilmark{4}}

\altaffiltext{1}{State Key Laboratory of Space Weather, National Space 
Science Center, Chinese Academy of Sciences, Beijing 100190, China;
liuxying@spaceweather.ac.cn}

\altaffiltext{2}{University of Chinese Academy of Sciences, No.19A 
Yuquan Road, Beijing 100049, China}

\altaffiltext{3}{Space Sciences Laboratory, University of California, Berkeley, 
CA 94720, USA}

\altaffiltext{4}{Kavli Institute for Astrophysics and Space Research, 
Massachusetts Institute of Technology, Cambridge, MA 02139, USA}

\begin{abstract}

The largest geomagnetic storms of solar cycle 24 so far occurred on 2015 March 17 and June 22 with $D_{\rm st}$ minima of $-223$ and $-195$ nT, respectively. Both of the geomagnetic storms show a multi-step development. We examine the plasma and magnetic field characteristics of the driving coronal mass ejections (CMEs) in connection with the development of the geomagnetic storms. A particular effort is to reconstruct the in situ structure using a Grad-Shafranov technique and compare the reconstruction results with solar observations, which gives a larger spatial perspective of the source conditions than one-dimensional in situ measurements. Key results are obtained concerning how the plasma and magnetic field characteristics of CMEs control the geomagnetic storm intensity and variability: (1) a sheath-ejecta-ejecta mechanism and a sheath-sheath-ejecta scenario are proposed for the multi-step development of the 2015 March 17 and June 22 geomagnetic storms, respectively; (2) two contrasting cases of how the CME flux-rope characteristics generate intense geomagnetic storms are found, which indicates that a southward flux-rope orientation is not a necessity for a strong geomagnetic storm; and (3) the unexpected 2015 March 17 intense geomagnetic storm resulted from the interaction between two successive CMEs plus the compression by a high-speed stream from behind, which is essentially the ``perfect storm" scenario proposed by \citet[][i.e., a combination of circumstances results in an event of unusual magnitude]{liu14a}, so the ``perfect storm" scenario may not be as rare as the phrase implies. 

\end{abstract}


\section{Introduction}

A topic of increasing interest to space weather is how the plasma and magnetic field characteristics of coronal mass ejections (CMEs) result in geomagnetic storm activity, in particular those intense events. The southward magnetic field and speed of CMEs at the Earth have received the most attention, because their cross product, the dawn-dusk electric field, controls the rate of the solar wind energy coupling to the terrestrial magnetosphere \citep{dungey61}. However, it is still not clear how the ejecta speed and southward magnetic field work together to achieve a sustained, enhanced dawn-dusk electric field and how they lead to the variability of geomagnetic storms. 

The southward magnetic field is often found within the ejecta reaching the Earth in the form of an interplanetary CME (ICME) with a preceding shock. This usually leads to a classic geomagnetic storm sequence: a sudden commencement generated by the shock, a main decrease phase caused by the ejecta's southward magnetic field, and then a recovery phase. In addition to the driver gas, the sheath region between the shock and ICME can also be geo-effective \citep[e.g.,][]{tsurutani88, liu08b} as both the sheath speed and southward magnetic field are amplified by shock compression. The sheath-ejecta scenario has been invoked to explain the two-step development of geomagnetic storms \citep{kamide98}. Complex ejecta resulting from interactions between CMEs \citep{burlaga01, burlaga02} can be very geo-effective owing to their prolonged durations \citep[e.g.,][]{farrugia04, zhang07, lugaz14, mishra15}. They could also cause two-step geomagnetic storms \citep{farrugia06, liu14b}. A special case of complex ejecta is the interaction of a preceding ejecta with an overtaking shock \citep[e.g.,][]{liu12, liu14b, mostl12, harrison12, webb13}. The shock enhances the pre-existing southward magnetic field inside the ejecta, an idea for increased geo-effectiveness dating back several decades \citep{burlaga91, vandas97}. A statistical analysis indicates that 19 out of 49 shocks propagating inside ICMEs are associated with an intense geomagnetic storm \citep[minimum $D_{\rm st} < -100$ nT;][]{lugaz15}. A recent study combining remote-sensing and in situ observations suggests a ``perfect storm" scenario for the generation of an extreme space weather event \citep{liu14a}: preconditioning of the upstream solar wind by an earlier CME plus in-transit interaction between later two closely launched CMEs, in order to have an exceptionally high solar wind speed and unusually strong ejecta magnetic fields at 1 AU. This, again, emphasizes the crucial importance of CME-CME interactions for space weather. 

On 2015 March 17 and June 22 the Earth underwent an intense geomagnetic storm with the minimum $D_{\rm st}$ of $-223$ and $-195$ nT, respectively. These are the largest geomagnetic storms of solar cycle 24 so far. They occurred in the declining phase of the solar cycle, a phenomenon that is not uncommon \citep{gopalswamy05, kilpua15}. We provide a timely analysis of the driving CMEs, in an attempt to identify the plasma and magnetic field characteristics controlling the geomagnetic storm intensity and variability. Another focus of this Letter is to test the ``perfect storm" scenario proposed by \citet{liu14a}: whether it is a rare coincidence or if it happens more frequently than what the phrase suggests. We examine the solar wind signatures and their connections with the development of the geomagnetic storms, complemented with the modeling of the $D_{\rm st}$ index using two empirical formulae based on the solar wind measurements \citep{burton75, om00}. We also use a Grad-Shafranov (GS) technique \citep{hau99, hu02}, which has been validated by well separated multi-spacecraft measurements \citep{liu08a, mostl09}, to reconstruct the in situ ICME structure. The GS method can give a cross section as well as flux-rope orientation without prescribing the geometry. In conjunction with solar observations it provides a larger spatial perspective of ICMEs than one-dimensional in situ measurements \citep{liu10}. These efforts are key to understanding how the plasma and magnetic field characteristics of CMEs are connected with the intensity of geomagnetic storms as well as their variability. 
 
\section{The 2015 March Event}

Tracing back to the Sun, the drivers of the 2015 March 17 geomagnetic storm were two interacting CMEs on March 15 (Figure~1, left). The second CME (CME2) had a maximum speed of about 1100 km s$^{-1}$ and was associated with a long-duration C9.1 flare from AR 12297 (S22$^{\circ}$W25$^{\circ}$) that peaked at 02:13 UT on March 15. The first CME (CME1) occurred on March 14 and had a speed of about 350 km s$^{-1}$. It was likely associated with a C2.6 flare from the same active region (S21$^{\circ}$W20$^{\circ}$) that peaked around 11:55 UT on March 14. A first impression from the coronagraph images is that CME1 was largely propagating southward while CME2 had a major component heading west. One may expect that the Earth would encounter the flank of the ejecta, so this would not raise the alarm for a major geomagnetic storm. Another fact that also makes the occurrence of an intense geomagnetic storm surprising is that the associated flares are relatively weak. In this sense, the occurrence of the 2015 March 17 intense geomagnetic storm is similar to the formation of the 2012 July 23 super solar storm that impacted STEREO A \citep{liu14a}. Without white-light observations from STEREO accurate CME kinematics cannot be obtained.   

Figure~2 shows the in situ signatures observed at Wind. A shock passed Wind at 04:01 UT on March 17 and caused the sudden commencement of the geomagnetic storm. It is difficult to unambiguously connect the in situ signatures with the coronagraph images without wide-angle imaging observations from STEREO. However, application of an empirical model \citep{gopalswamy00} with the CME speed of 1100 km s$^{-1}$ gives a predicted arrival time of 23:59 UT on March 16 at Wind (0.99 AU from the Sun), which is only 4 hours earlier than the observed shock arrival. As can be seen from the figure, two ICMEs (or flux ropes) are identified. Our interpretation is different from those of \citet{kataoka15} and \citet{gopalswamy15} who identify a single, shorter ICME interval from the data (although with different durations). The reason that we believe there are two ICMEs is as follows. First, there are multiple rotations in the magnetic field components whose polarities change twice in the shaded data intervals. Obviously these features cannot be explained by a single flux rope. Second, our ICME intervals are an outcome of the GS reconstruction, which is sensitive to the chosen boundaries. Despite the magnetic field fluctuations, both of the ICMEs can be reconstructed fairly well (see description below). Third, the interpretation of two ICMEs is consistent with what the solar observations indicate (see Figure~1 and discussions below). One may argue for a single ICME interval (say, from 12:58 UT on March 17 to 03:22 UT on March 18) based on the depressed proton temperature, a signature often used to identify ICMEs. A reasonable GS reconstruction, however, cannot be obtained for the interval and its variations. Given the presence of the compression by a high-speed stream from behind and CME-CME interactions, the criterion of a low proton temperature for identifying ICMEs may not be valid in the current case. 

Within the ICME intervals the maximum magnetic field strength is about 33 nT while the southward component reaches $-25$ nT. These are not small magnetic fields considering only the flanks of the CMEs are encountered (see Figure~1). Interactions between these two ICMEs may have inhibited their expansion, a mechanism to create strong ejecta magnetic fields as we have seen from the 2012 July 23 event \citep{liu14a}. Also note the high-speed stream compressing ICME2 from behind, which may help maintain a strong ejecta magnetic field and a relatively high speed as well. The $D_{\rm st}$ profile indicates a two-step geomagnetic storm sequence with a global minimum of $-223$ nT. The first dip is produced by the southward magnetic field component in the sheath region behind the shock, while the second one results from the southward fields within the two ICMEs that last about $12$ hours. Given the presence of a preceding shock and two interacting ICMEs, the two-step development of the geomagnetic storm can be classified as a sheath-ejecta-ejecta scenario. The modeled $D_{\rm st}$ index using the \citet{om00} formula (minimum $-170$ nT) generally agrees with actual $D_{\rm st}$ measurements but underestimates the global minimum. The \citet{burton75} scheme gives a deeper global minimum ($-280$ nT) and a shallower recovery phase than measured.

Further information on how the plasma and magnetic field characteristics control the geomagnetic storm activity is obtained from the GS reconstruction, as shown in Figure~3. The reconstructions give a right-handed flux rope structure for both of the ICMEs, as can be judged from the transverse fields along the spacecraft trajectory together with the axial direction. Their axis orientations are almost opposite to each other: an elevation angle of about $33^{\circ}$ and azimuthal angle of about $256^{\circ}$ (in RTN coordinates) for ICME1, and an elevation angle of about $-18^{\circ}$ and azimuthal angle of about $92^{\circ}$ for ICME2. These low inclinations are consistent with the slightly tilted neutral line (N. Gopalswamy, private communication) and filament channel (M. Temmer, private communication) associated with the active region. ICME1 has a larger elevation angle ($33^{\circ}$), which may help explain the encounter in spite of the largely southward propagation direction of CME1 (Figure~1). The impact of ICME2 may be accounted for by its lower elevation angle ($-18^{\circ}$), although CME2 had a major section propagating westward. In addition, the angular span of an ICME can be $60^{\circ}$ or even larger \citep{bothmer98, richardson02}, so it is not surprising that both of the ICMEs hit the Earth. It is also likely that the CME-CME interactions, the following high-speed stream and even the surrounding coronal magnetic field structures may have changed the propagation directions of both CMEs \citep[e.g.,][]{gopalswamy09, zuccarello12, lugaz12, liu14a, mostl15, kay15}.

Given those relatively low inclination angles, the geomagnetic storm was mainly caused by the azimuthal magnetic field components of the flux ropes rather the axial components. A particularly interesting feature shown by the cross section of ICME1 is that the azimuthal field component (with a maximum value of about 30 nT) is much larger than the axial component (maximum value of only 8 nT). Therefore, if the flux rope were vertically orientated with the axis pointing southward the geomagnetic storm would be much weaker since the southward field would be less. This is contrary to the common belief that a southward pointing flux rope favors a strong geomagnetic storm. ICME2 has comparable axial and azimuthal magnetic field components. Another prominent feature is the vortices visible in the cross section of ICME2, reminiscent of the Kelvin-Helmholtz instability. This is probably a result of the fast stream interacting with ICME2 from behind (see Figure~2). 

\section{The 2015 June Event}

In 2015 June AR 12371 exhibited elevated activity, somewhat similar to AR 11429 in 2012 March \citep{liu13, liu14c, wang14, sun15}. The active region produced a CME of about 1200 km s$^{-1}$ associated with an M3.0 flare from N13$^{\circ}$E45$^{\circ}$ that peaked at 17:36 UT on June 18, a CME of about 1300 km s$^{-1}$ associated with an M2.0 flare from N12$^{\circ}$E13$^{\circ}$ peaking at 01:42 UT on June 21, a CME of about 1000 km s$^{-1}$ associated with an M6.5 flare from N13$^{\circ}$W05$^{\circ}$ peaking at 18:23 UT on June 22, and another one of about 1700 km s$^{-1}$ associated with an M7.9 flare from N10$^{\circ}$W42$^{\circ}$ peaking at 08:16 UT on June 25. All these CMEs impacted the Earth. The June 21 CME appeared as a single halo event in the coronagraph images (see Figure~1, right), and a near head-on collision with the Earth was expected. 

The corresponding in situ signatures at Wind are displayed in Figure~4. A cluster of shocks passed Wind at 16:05 UT on June 21, 05:02 UT and 18:00 UT on June 22, and 13:12 UT on June 24, respectively. The ICME boundaries are determined from the magnetic field in conjunction with the proton temperature and density. The first shock (S1) seemed driven by the June 18 CME, and the second one (S2) was likely associated with a CME from June 19. No driver signatures are observed at Wind for these two shocks, presumably owing to the largely eastward and southward propagation directions of the June 18 and 19 CMEs respectively (not shown here). The ICME and its preceding shock (S3) were produced by the June 21 CME; again, application of the empirical model \citep{gopalswamy00} with the CME speed of 1300 km s$^{-1}$ yields a predicted arrival time of 17:02 UT on June 22 at Wind (1.02 AU from the Sun), which is only 1 hour earlier than the observed S3 arrival. The fourth shock (S4) that was overtaking the ICME at 1 AU was associated with the June 22 CME. A series of dips in the magnetic field strength are observed inside the ICME, suggestive of the presence of current sheets. This signature is possibly due to the heliospheric current sheet cutting through the ejecta, which may lead to a chain of small flux ropes within the ICME (see below). 

The $D_{\rm st}$ profile shows a multi-step geomagnetic storm with a global minimum of $-195$ nT. The first dip is produced by the fluctuating southward field component upstream of S3 (likely owing to amplification by the two preceding shocks), the second one by the southward field in the sheath downstream of S3 (further enhanced by S3), and the major dip by the southward field in the first hatched interval inside the ejecta. The southward field in the second hatched interval only creates a flattening of the $D_{\rm st}$ value, perhaps because of the extremely low density. Note that the solar wind density upstream of the ejecta is significantly enhanced by the three preceding shocks. This high density may feed the plasma sheet of the magnetosphere, which in turn helps intensify the ring current \citep{farrugia06, lavraud06}. Given the presence of more than one preceding shocks and a single ICME, the multi-step development of the geomagnetic storm can be classified as a sheath-sheath-ejecta scenario. Again, the modeled $D_{\rm st}$ index using the \citet{om00} formula (minimum $-174$ nT) underestimates the global minimum but reproduces the recovery phase fairly well, whereas application of the \citet{burton75} scheme gives a larger global minimum ($-241$ nT) than measured. 

Figure~5 shows the reconstructed cross sections of two small flux ropes identified inside the ICME. Here we call them small flux ropes rather than ICMEs, in order to distinguish from the 2015 March 17 case. Specifically, the June 22 event is a single ejecta instead of multiple ICMEs. These small flux ropes may have formed from the interaction between the CME and the heliospheric current sheet. The reconstructions yield a left-handed structure for both of the flux ropes and similar axis orientations: an elevation angle of about $-77^{\circ}$ and azimuthal angle of about $236^{\circ}$ for the first one (FR1), and an elevation angle of about $-61^{\circ}$ and azimuthal angle of about $272^{\circ}$ for the second one (FR2). The same chirality and similar axis orientations support the interpretation of a single ICME. Both flux ropes have strong axial magnetic field components compared with the azimuthal ones. Also note the largely southward orientation of the flux ropes. It is the strong axial field component on top of the largely southward flux-rope orientation, in conjunction with the relatively high solar wind speed, that may have resulted in the intense geomagnetic storm. 

\section{Conclusions}

We have examined the sources of the 2015 March 17 and June 22 intense geomagnetic storms, the largest ones of solar cycle 24 up to the time of this writing. Key findings are obtained on how the plasma and magnetic field characteristics of CMEs control the geomagnetic storm intensity and variability: 

1. A sheath-ejecta-ejecta mechanism and a sheath-sheath-ejecta scenario are proposed for the development of multi-step geomagnetic storms, based on the observed and reconstructed solar wind structures associated with the CMEs. The 2015 March 17 geomagnetic storm shows a two-step development, which is produced by the southward magnetic field components behind the preceding shock and those within two interacting CMEs. This falls into the sheath-ejecta-ejecta category. The 2015 June 22 geomagnetic storm exhibits a multi-step development, which is caused by the southward fields due to amplification by a series of preceding shocks and those within a single ejecta. This is classified as a sheath-sheath-ejecta scenario. The multiple preceding shocks and sheaths may precondition the magnetosphere for the growth of an intense geomagnetic storm. 

2. We find two contrasting cases of how the CME flux-rope characteristics generate intense geomagnetic storms. Our GS reconstruction of the ejecta responsible for the 2015 June 22 geomagnetic storm indicates that the geomagnetic storm resulted from the largely southward flux-rope orientation with a strong axial magnetic field component. However, for the 2015 March 17 geomagnetic storm the GS reconstruction reveals a much larger azimuthal field component than the axial component. The intense geomagnetic storm occurred despite low flux-rope inclinations. A southward flux-rope orientation is thus not a necessity for a strong geomagnetic storm to occur. 

3. The ``perfect storm" scenario proposed by \citet{liu14a} may not be as rare as the phrase implies. The 2015 March 17 intense geomagnetic storm occurred in spite of the relatively weak solar flares and an encounter with the CME flank. What makes it an intense geomagnetic storm is the interaction between two successive CMEs plus the compression by a high-speed stream from behind, which helps maintain strong ejecta magnetic fields and a relatively high speed. This is essentially the ``perfect storm" scenario -- a combination of circumstances results in an event of unusual magnitude, although the 2015 March 17 event is not ``super" in the same sense as the 2012 July 23 solar storm. Note that there are many combinations of circumstances that can occur to make an event more geo-effective, including pileup of events, pre-event rarefactions and field line stretching, shock enhancement of southward fields, and following high-speed streams causing compressions. This ``perfect storm" scenario now seems useful and necessary to worry about because complex events with these combinations are common. 

\acknowledgments The research was supported by the Recruitment Program of Global Experts of China, NSFC under grant 41374173 and the Specialized Research Fund for State Key Laboratories of China. We acknowledge the use of data from Wind and SOHO and the $D_{\rm st}$ index from WDC in Kyoto.

\clearpage

\begin{figure}
\epsscale{0.9} \plotone{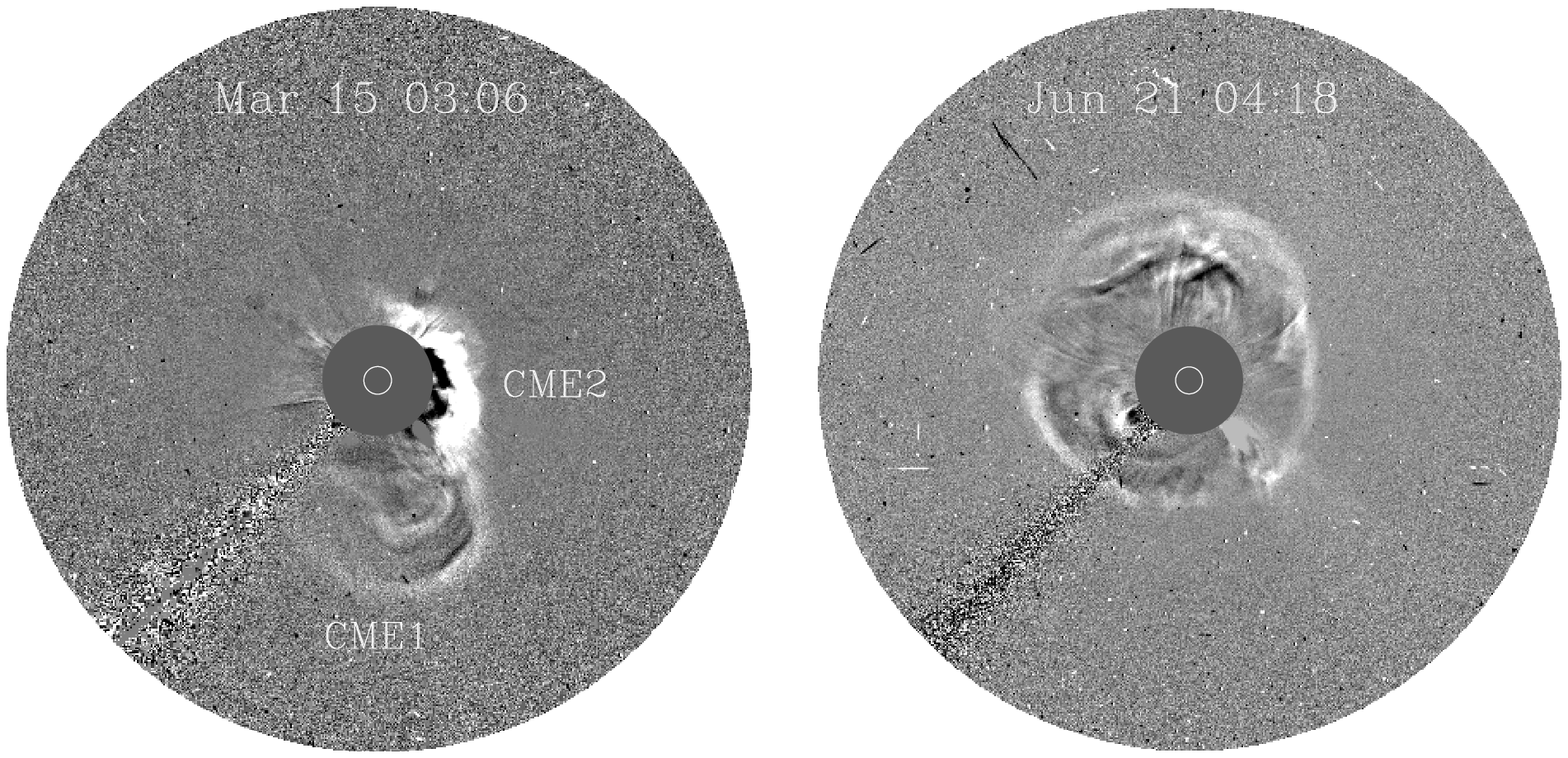} 
\caption{Difference images of the 2015 March 15 (left) and June 21 (right) CMEs from LASCO C3 aboard SOHO. Note the interaction between the March 15 CME (CME2) and a preceding one that occurred on March 14 (CME1).}
\end{figure}

\clearpage

\begin{figure}
\epsscale{0.75} \plotone{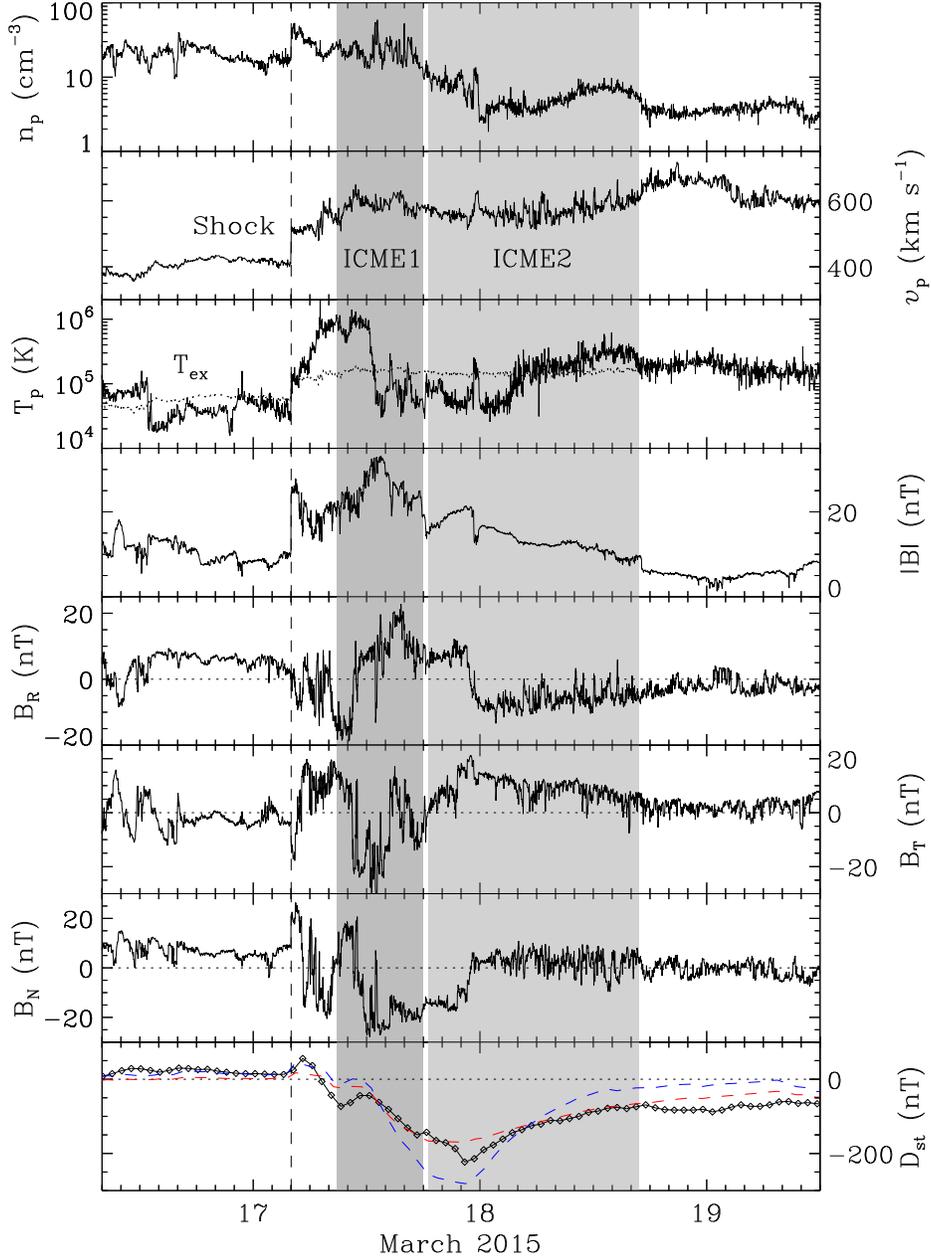} 
\caption{Solar wind measurements at Wind and associated $D_{\rm st}$ index for the 2015 March 17 event. From top to bottom, the panels show the proton density, bulk speed, proton temperature, magnetic field strength and components, and $D_{\rm st}$ index, respectively. The shaded regions indicate two ICME intervals. The vertical dashed line marks the associated shock. The dotted curve in the third panel denotes the expected proton temperature calculated from the observed speed \citep{lopez87}. The red and blue curves in the bottom panel represent $D_{\rm st}$ values estimated using the formulae of \citet{om00} and \citet{burton75}, respectively.}
\end{figure}

\clearpage

\begin{figure}
\centerline{\includegraphics[width=36pc]{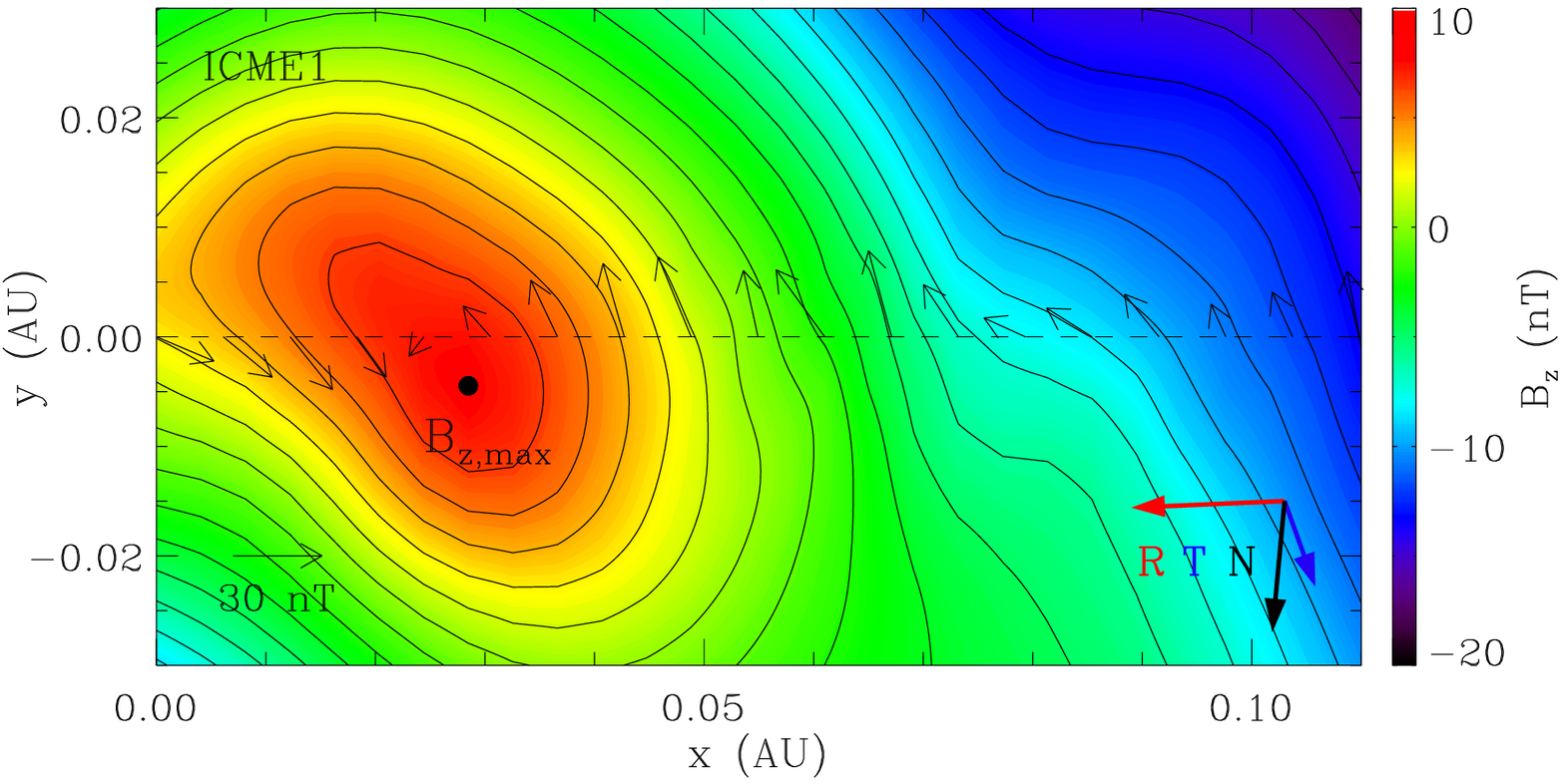}}
\vspace{0.5pc}
\centerline{\includegraphics[width=36pc]{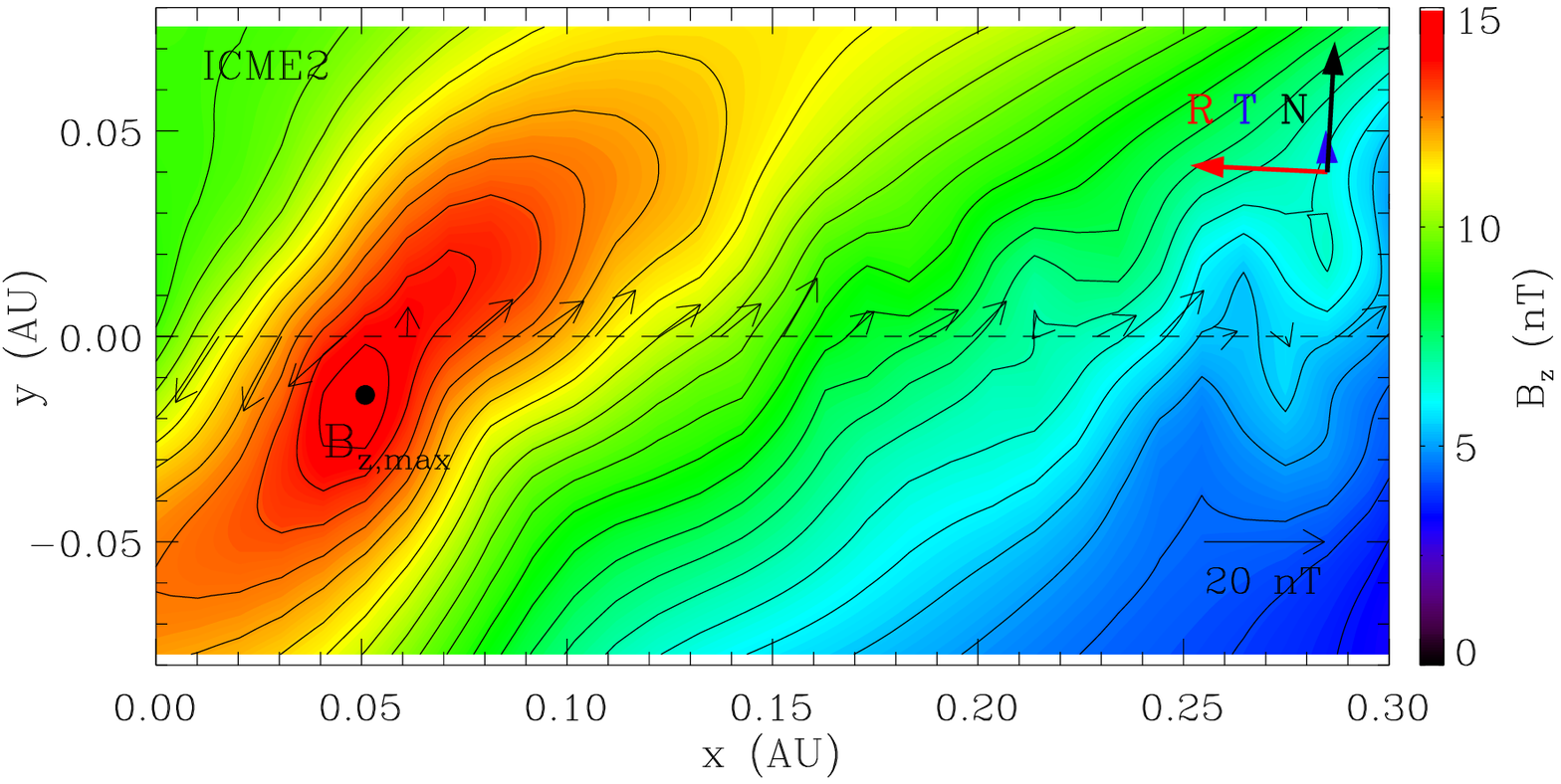}}
\caption{Reconstructed cross sections of ICME1 (upper) and ICME2 (lower). Black contours show the distribution of the vector potential, and the color shading indicates the value of the axial magnetic field. The location of the maximum axial field is indicated by the black dot. The dashed line marks the trajectory of the Wind spacecraft. The thin black arrows denote the direction and magnitude of the observed magnetic fields projected onto the cross section, and the thick colored arrows show the projected RTN directions.}
\end{figure}

\clearpage

\begin{figure}
\epsscale{0.75} \plotone{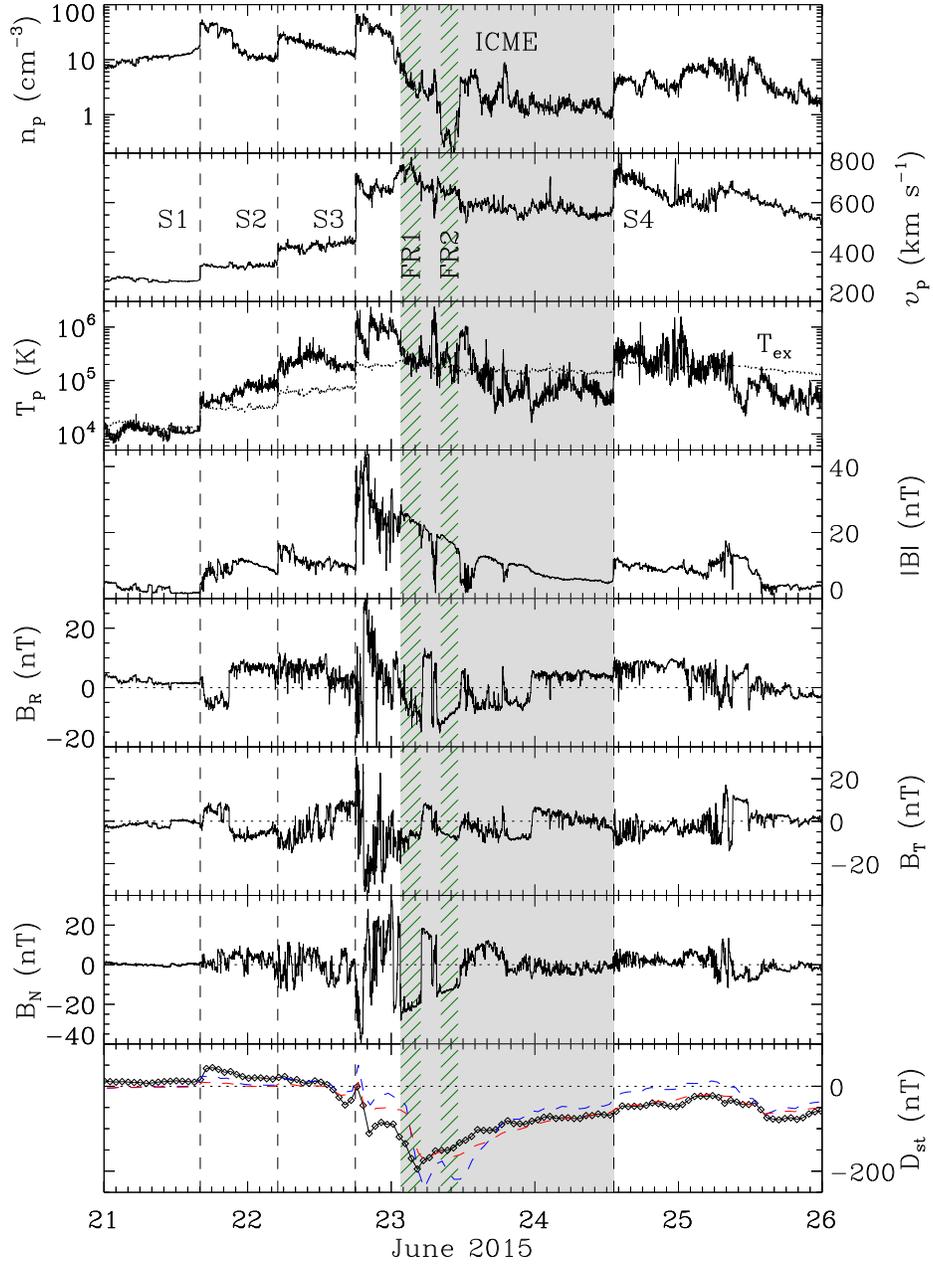} 
\caption{Solar wind measurements at Wind and associated $D_{\rm st}$ index for the 2015 June 22 event. Similar to Figure~2. The shaded region shows the overall ejecta interval, while the hatched areas indicate two small flux ropes identified within the ICME. Three shocks are observed ahead of the ejecta. The last shock (S4) was driven by the CME that occurred at the Sun on 2015 June 22 and was overtaking the ICME at 1 AU.}
\end{figure}

\clearpage

\begin{figure}
\centerline{\includegraphics[width=36pc]{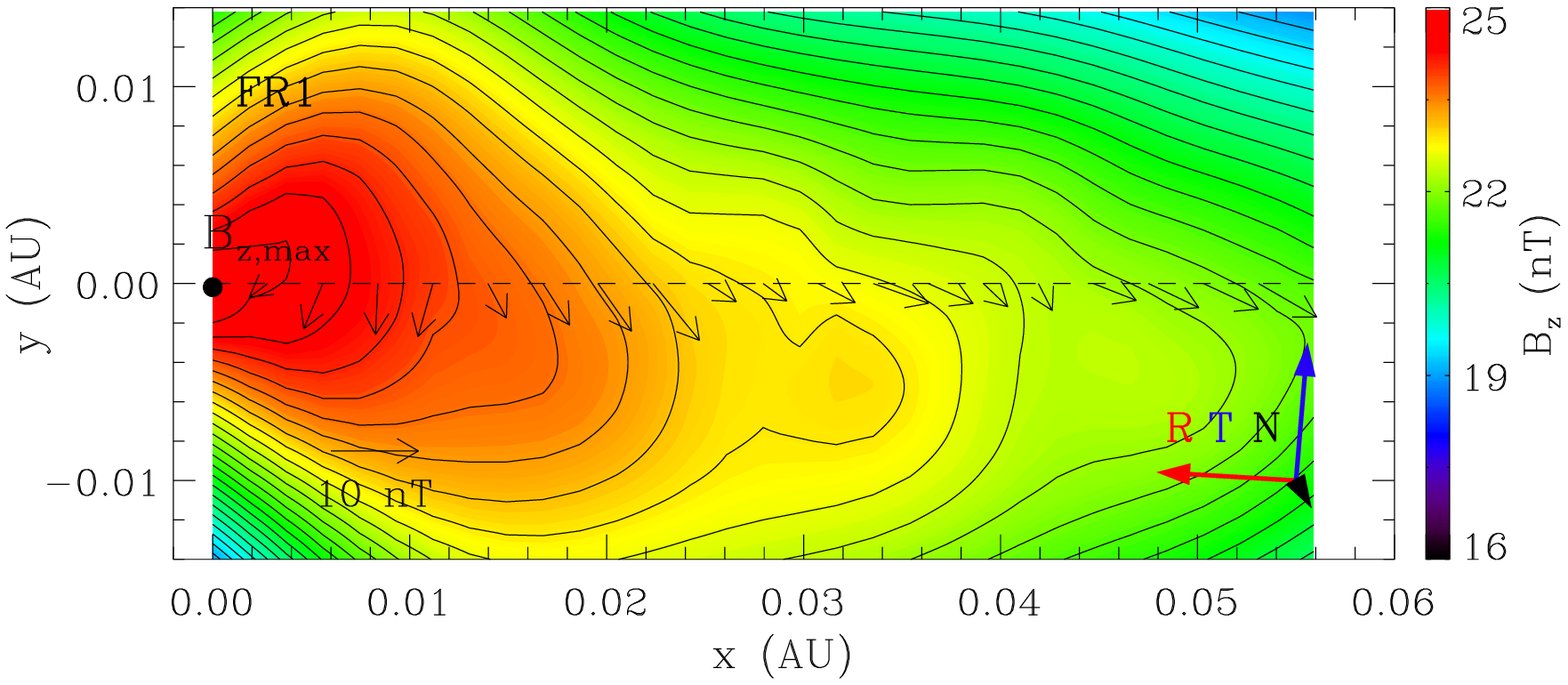}}
\vspace{0.5pc}
\centerline{\includegraphics[width=36pc]{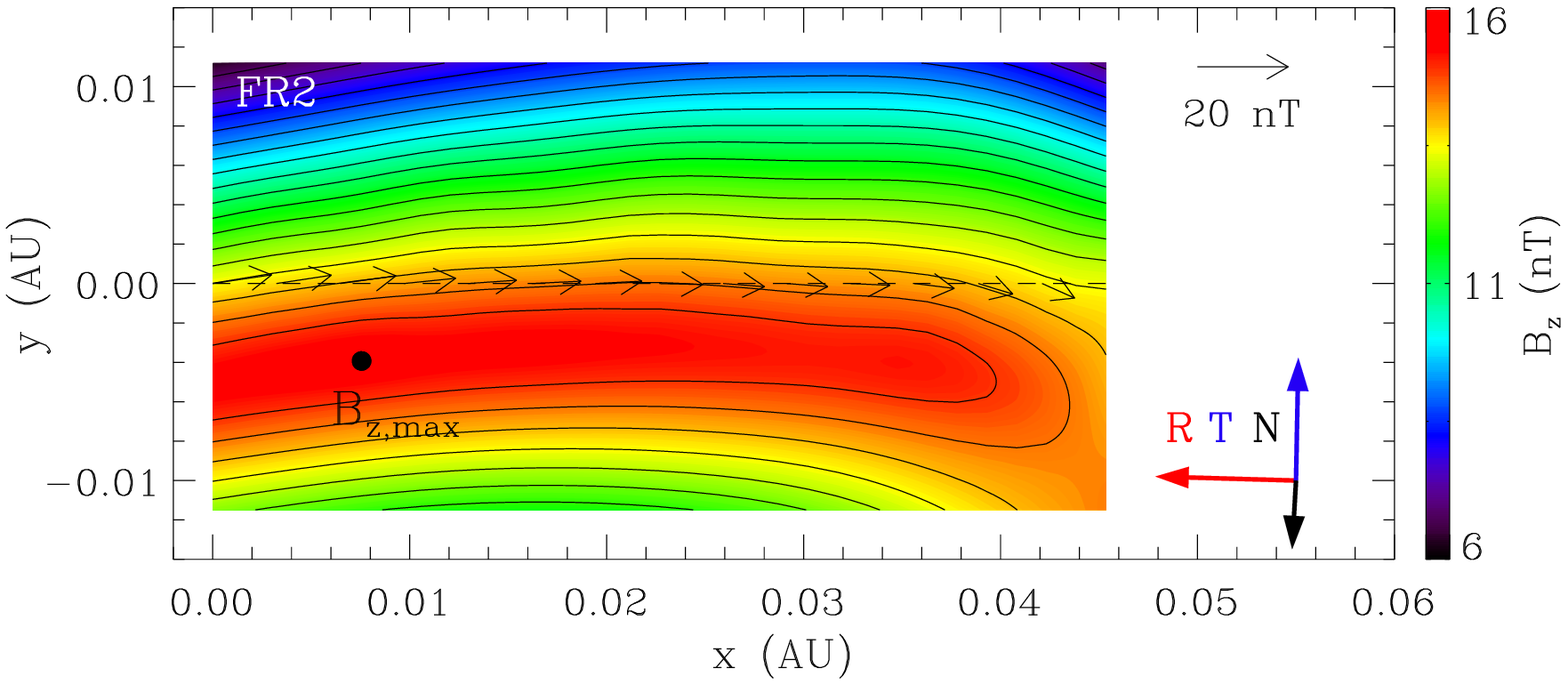}}
\caption{Reconstructed cross sections of FR1 (upper) and FR2 (lower) identified inside the 2015 June 22 ICME. Similar to Figure~3. These small flux ropes are both left-handed and have similar axis orientations.}
\end{figure}


\begin{thebibliography}{}

\bibitem[Bothmer \& Schwenn(1998)]{bothmer98}
Bothmer, V., \& Schwenn, R. 1998, Ann. Geophys., 16, 1

\bibitem[Burlaga(1991)]{burlaga91}
Burlaga, L. F. 1991, in Physics of the Inner Heliosphere II: 
Particles, Waves and Turbulence, ed. R. Schwenn, E. Marsch (Springer, NY), 1

\bibitem[Burlaga et al.(2001)]{burlaga01}
Burlaga, L. F., Skoug, R. M., Smith, C. W., Webb, D. F., Zurbuchen, T. H., \& Reinard, A. 
2001, \jgr, 106, 20957 

\bibitem[Burlaga et al.(2002)]{burlaga02}
Burlaga, L. F., Plunkett, S. P., \& St. Cyr, O. C. 2002, \jgr, 107,
1266

\bibitem[Burton et al.(1975)]{burton75}
Burton, R. K., McPherron, R. L., \& Russell, C. T. 1975, \jgr, 80, 4204

\bibitem[Dungey(1961)]{dungey61}
Dungey, J. W. 1961, \prl, 6, 47

\bibitem[Farrugia \& Berdichevsky(2004)]{farrugia04}
Farrugia, C., \& Berdichevsky, D. 2004, Ann. Geophys., 22, 3679

\bibitem[Farrugia et al.(2006)]{farrugia06}
Farrugia, C. J., Jordanova, V. K., Thomsen, M. F., Lu, G., Cowley, S. W. H., \&
Ogilvie, K. W. 2006, \jgr, 111, A11104

\bibitem[Gopalswamy et al.(2000)]{gopalswamy00}
Gopalswamy, N., Lara, A., Lepping, R. P., Kaiser, M. L., Berdichevsky, D., \& 
St. Cyr, O. C. 2000, \grl, 27, 145

\bibitem[Gopalswamy et al.(2005)]{gopalswamy05}
Gopalswamy, N., et al. 2005, \jgr, 110, A09S15

\bibitem[Gopalswamy et al.(2009)]{gopalswamy09}
Gopalswamy, N., M\"akel\"a, P., Xie, H., Akiyama, S., \& Yashiro, S. 2009, 
\jgr, 114, A00A22 

\bibitem[Gopalswamy et al.(2015)]{gopalswamy15}
Gopalswamy, N., et al. 2015, in Proc. 14th International Ionospheric 
Effects Symposium (Alexandria, VA), in press

\bibitem[Harrison et al.(2012)]{harrison12}
Harrison, R. A., et al. 2012, \apj, 750, 45

\bibitem[Hau \& Sonnerup(1999)]{hau99}
Hau, L.-N., \& Sonnerup, B. U. \"{O}. 1999, \jgr, 104, 6899

\bibitem[Hu \& Sonnerup(2002)]{hu02}
Hu, Q., \& Sonnerup, B. U. \"{O}. 2002, \jgr, 107, 1142

\bibitem[Kamide et al.(1998)]{kamide98}
Kamide, Y., et al. 1998, \jgr, 103, 6917

\bibitem[Kataoka et al.(2015)]{kataoka15}
Kataoka, R., Shiota, D., Kilpua, E., \& Keika, K. 2015, \grl, 
doi: 10.1002/2015GL064816

\bibitem[Kay et al.(2015)]{kay15}
Kay, C., Opher, M., \& Evans, R. M. 2015, \apj, 805, 168

\bibitem[Kilpua et al.(2015)]{kilpua15}
Kilpua, E. K. J., et al. 2015, \apj, 806, 272

\bibitem[Lavraud et al.(2006)]{lavraud06}
Lavraud, B., Thomsen, M. F., Borovsky, J. E., Denton, M. H., \& 
Pulkkinen, T. I. 2006, \jgr, 111, A09208

\bibitem[Liu et al.(2008a)]{liu08a}
Liu, Y., et al. 2008a, \apj, 677, L133

\bibitem[Liu et al.(2008b)]{liu08b}
Liu, Y., Manchester, W. B., Richardson, J. D., Luhmann, J. G., Lin, R. P., 
\& Bale, S. D. 2008b, \jgr, 113, A00B03

\bibitem[Liu et al.(2010)]{liu10}
Liu, Y., Thernisien, A., Luhmann, J. G., Vourlidas, A., Davies, J.
A., Lin, R. P., \& Bale, S. D. 2010, \apj, 722, 1762

\bibitem[Liu et al.(2012)]{liu12}
Liu, Y. D., et al. 2012, \apj, 746, L15

\bibitem[Liu et al.(2013)]{liu13}
Liu, Y. D., et al. 2013, \apj, 769, 45

\bibitem[Liu et al.(2014a)]{liu14a}
Liu, Y. D., et al. 2014a, Nat. Commun., 5, 3481, doi: 10.1038/ncomms4481

\bibitem[Liu et al.(2014b)]{liu14b}
Liu, Y. D., et al. 2014b, \apj, 793, L41

\bibitem[Liu et al.(2014c)]{liu14c}
Liu, Y. D., Richardson, J. D., Wang, C., \& Luhmann, J. G. 2014c, \apj, 788, L28

\bibitem[Lugaz et al.(2012)]{lugaz12}
Lugaz, N., et al. 2012, \apj, 759, 68

\bibitem[Lugaz \& Farrugia(2014)]{lugaz14}
Lugaz, N., \& Farrugia, C. J. 2014, \grl, 41, 769

\bibitem[Lugaz et al.(2015)]{lugaz15}
Lugaz, N., Farrugia, C. J., Smith, C. W., \& Paulson, K. 2015, 
\jgr, 120, 2409

\bibitem[Lopez(1987)]{lopez87}
Lopez, R. E. 1987, \jgr, 92, 11189

\bibitem[Mishra et al.(2015)]{mishra15}
Mishra, W., Srivastava, N., \& Chakrabarty, D. 2015,  
\solphys, 290, 527

\bibitem[M\"{o}stl et al.(2009)]{mostl09}
M\"{o}stl, C., et al. 2009, \jgr, 114, A04102

\bibitem[M\"{o}stl et al.(2012)]{mostl12}
M\"{o}stl, C., et al. 2012, \apj, 758, 10

\bibitem[M\"{o}stl et al.(2015)]{mostl15}
M\"{o}stl, C., et al. 2015, Nat. Commun., 6, 7135, doi: 10.1038/ncomms8135

\bibitem[O'Brien \& McPherron(2000)]{om00}
O'Brien, T. P., \& McPherron, R. L. 2000, \jgr, 105, 7707

\bibitem[Richardson et al.(2002)]{richardson02}
Richardson, J. D., Paularena, K. I., Wang, C., \& Burlaga, L. F. 2002, 
\jgr, 107, 1041

\bibitem[Sun et al.(2015)]{sun15}
Sun, X., et al. 2015, \apj, 804, L28

\bibitem[Tsurutani et al.(1988)]{tsurutani88}
Tsurutani, B. T., Smith, E. J., Gonzalez, W. D., Tang, F., \& Akasofu, S. I. 
1988, \jgr, 93, 8519

\bibitem[Vandas et al.(1997)]{vandas97}
Vandas, M., Fischer, S., Dryer, M., Smith, Z., Detman, T., \&
Geranios, A. 1997, \jgr, 102, 22295

\bibitem[Wang et al.(2014)]{wang14}
Wang, R., Liu, Y. D., Yang, Z., \& Hu, H. 2014, \apj, 791, 84

\bibitem[Webb et al.(2013)]{webb13}
Webb, D. F., et al. 2013, \solphys, 285, 317

\bibitem[Zhang et al.(2007)]{zhang07}
Zhang, J., et al. 2007, \jgr, 112, A10102

\bibitem[Zuccarello et al.(2012)]{zuccarello12}
Zuccarello, F. P., Bemporad, A., Jacobs, C., Mierla, M., Poedts, S., \& 
Zuccarello, F. 2012, \apj, 744, 66

\end{thebibliography}
\end{document}